\documentclass[twoside,11pt]{article}
\usepackage{jair}
\usepackage{rawfonts}

\usepackage{url}
\usepackage{multirow,makecell}
\usepackage{amsmath,amssymb,amsthm}
\usepackage{algorithmic,algorithm}
\usepackage{mathtools}
\usepackage{bbm}
\usepackage{enumerate}
\usepackage[round]{natbib}

\newtheorem{thm}{Theorem}
\newtheorem{defn}{Definition}
\theoremstyle{definition}
\newtheorem*{remark}{Remark}
\newtheorem{remark-star}{Remark}
\newtheorem{remark-star-1}{Remark}

\newenvironment{proof-sketch}{%
  \proof}{\endproof}

\def\R{\mathbb{R}}
\def\E{\mathbb{E}}
\def\P{\mathbb{P}}

\def\Var{\mathrm{Var}}

\jairheading{73}{2022}{209-229}{09/2021}{01/2022}
\ShortHeadings{Doubly Robust Crowdsourcing}
{Liu \& Wang}
\firstpageno{209}

\begin{document}

\title{Doubly Robust Crowdsourcing}

\author{\name Chong Liu \email chongliu@cs.ucsb.edu \\
       \name Yu-Xiang Wang \email yuxiangw@cs.ucsb.edu \\
       \addr Department of Computer Science\\
       University of California, Santa Barbara\\
       Santa Barbara, CA 93106, USA}


\maketitle

\begin{abstract}
Large-scale \emph{labeled} dataset is the indispensable fuel that ignites the AI revolution as we see today. Most such datasets are constructed using crowdsourcing services such as Amazon Mechanical Turk which provides noisy labels from non-experts at a fair price. The sheer size of such datasets mandates that it is only feasible to collect a few labels per data point. We formulate the problem of test-time label aggregation as a statistical estimation problem of inferring the expected voting score. By imitating workers with supervised learners and using them in a doubly robust estimation framework, we prove that the variance of estimation can be substantially reduced, even if the learner is a poor approximation. Synthetic and real-world experiments show that by combining the doubly robust approach with adaptive worker/item selection rules, we often need much lower label cost to achieve nearly the same accuracy as in the ideal world where all workers label all data points. 
\end{abstract}

\section{Introduction}\label{sec:intro}

The rise of machine learning approaches in artificial intelligence has enabled machines to perform well on many cognitive tasks that were previously thought of as what makes us human. In many specialized tasks, for example, animial recognition in images \citep{he2015delving}, conversational speech recognition \citep{xiong2018microsoft}, translating Chinese text into English \citep{hassan2018achieving}, learning-based systems are shown to have reached and even surpassed human-level performances. These remarkable achievements could not have been possible without the many large-scale datasets that are made available by researchers over the past two decades. ImageNet, for instance, has long been regarded as what spawned the AI revolution that we are experiencing today. These labels do not come for free. ImageNet's 11 million images were labeled using Amazon Mechanical Turk (AMT) into more than 15,000 synsets (classes in an ontology). On average, each image required roughly $2-5$ independent human annotations, which were provided by 25,000 AMT workers over a period of three years.
We estimate that the cost of getting all these annotations goes well above one million dollars.

As the deep learning models get larger and more powerful every day so as to tackle some of the more challenging AI tasks, their ferocious appetites for even larger labeled dataset have grown tremendously as well.
However, unlike the abundant unlabeled data, it is often difficult, expensive, or even impossible to consult expert opinions on large number of items. Here the items can be images, documents, voices, sentences, and so on.
Services such as AMT have made it much easier to seek the wisdom of the crowd by having non-experts (called workers in the remainder of this paper) to provide many noisy annotations at a much lower cost.
A large body of work has been devoted to finding more scalable solutions. These include a variety of label-aggregation methods \citep{sheng2008get,welinder2010multidimensional,zhang2016spectral,zhou2015regularized}, end-to-end human-in-the-loop learning \citep{khetan2017learning}, online/adaptive worker selections \citep{branson2017lean,van2018lean} and so on.
At the heart of these approaches, there are various ways to evaluate individual worker performances and quantify the uncertainty in their provided labels.

In this paper, we take a pre-trained crowdsourcing model with worker evaluation as a blackbox and consider the problem of true label inference for new data points. We formulate this problem as a statistical estimation problem and propose a number of ways to radically reduce the number of worker annotations.
\begin{enumerate}
    \item \textbf{Worker imitation} We propose to imitate each worker with a simple supervised learner that learns to predict the worker's label using the item feature.
    \item \textbf{Doubly robust crowdsourcing (DRC)} By tapping into the literature on doubly robust estimation, we design algorithms that exploit the possibly unreliable imitation agents and significantly reduce the estimation variance (hence annotation cost) while remaining unbiased.
    \item \textbf{Adaptive worker/item selection (AWS/AIS)} We propose to bootstrap the imitation agents' confidence estimates to adaptively filter out high confidence items and select the most qualified workers for low-confidence item, without additional cost.
\end{enumerate}

Our results are summarized as follows.
\begin{enumerate}
    \item We theoretically show that DRC technique can be used to generically improve any given crowdsourcing models using any nontrivial learned imitation agents.
    \item Synthetic and real-world experiments show DRC improves the label accuracy over the standard probabilistic inference with Dawid-Skene and majority voting models in almost all budget levels and all datasets.
    \item AWS and AIS often reduce the cost by orders of magnitudes, while enjoying the same level of accuracy. On several datasets, the proposed technique can often get away with much fewer annotations per item while achieving almost the same accuracy that can be obtained by having all workers annotating all items.
\end{enumerate}

\section{Related Work}\label{sec:related}
In this section, we briefly summarize the related work.

Our study is motivated by the many trailblazing approaches in label-aggregation including the wisdom-of-crowds \citep{welinder2010multidimensional}, Dawid-Skene model \citep{dawid1979maximum,zhang2016spectral}, minimax entropy approach \citep{zhou2015regularized}, permutation-based model \citep{shah2020permutation}, worker cluster model \citep{imamura2018anlysis}, crowdsourced regression model \citep{ok19a} and so on. Our contribution is complementary as we can take any of these models as blackboxes and hopefully improve their true-label inference. 

Doubly robust techniques originates from the causal inference literature \citep{rotnitzky1995semiparametric,bang2005doubly} and the use of it for variance reduction had led to several breakthroughs in machine learning \citep[e.g.,][]{johnson2013accelerating,wang2013variance}. We drew our inspirations directly from the use of doubly robust techniques in the off-policy evaluation problem in bandits and reinforcement learning \citep{dudik2014doubly,jiang2016doubly,wang2017optimal}. The variance analysis and weight-clipping are adapted from the calculations in \citet{dudik2014doubly} and \citet{wang2017optimal} with some minor differences. To the best of our knowledge, this is the first paper considering doubly robust techniques in crowdsourcing.

Our idea of adaptive item/worker selection is inspired by the recent work of \citet{branson2017lean} and \citet{van2018lean}. They propose an AI-aided approach that reduces the number of worker labels per item to be smaller than $1$ in an object detection task. The key idea is to train a computer vision algorithm to detect the bounding boxes using the aggregated labels that have been obtained thus far and if the algorithm achieves a high confidence on a new image, then the annotation provided by the algorithm is taken.

The differences of our work is twofold. First, our use of supervised learner is not to predict the true labels but rather to imitate workers. Second, our confidence measure is determined by supervised learners' approximation to what all workers would say about an item, rather than as a prior distribution added to model-based probabilistic inference.

\section{Problem Setup}\label{sec:problem}
In this section, we introduce the notations and formulate the problem as a statistical estimation problem.

\subsection{Notations}
Suppose we have $n$ items, $m$ workers, and $k$ classes. We adopt the notation $[k]:=\{1,2,3,...,k\}$.
Each item $j\in [n]$ is described as a $d$-dimensional feature vector $x_j$, and the feature matrix is $X=[x_1, x_2, \cdots, x_n]^\top \in \R^{n\times d}$. Each item $j\in [n]$ also has a hidden true label $y_j\in [k]$ which indicates the correct class that item $j$ belongs to.

Workers, such as those on AMT, are requested to classify items into one of the $k$ classes. We denote the label that worker $i\in[m]$ assigns to item $j$ as $\ell_{ij} \in [k]$. It is important to distinguish the worker-produced labels $\ell_{ij}$ with the true label $y_j$, as the workers are considered non-experts and they make mistakes. From here onwards, we will refer to the potentially noisy and erroneous labels from workers as ``annotations''. Conveniently, we also collect $\ell_{ij}$ into a matrix $L\in ([k]\cup\{\perp\})^{m\times n}$, where any entries in $L$ that are $\perp$ are unobserved labels. We use $\Omega\subset [m]\times[n],\Omega_i\subset [n], \Omega_j \subset[m]$ to denote the indices of the observed annotations, indices of all items worker $i$ annotated and indices of all workers that annotated item $j$ respectively. 
For a generic item $(x,y)$, $\Omega_x$ collects the indices of workers who annotated the item and the corresponding annotation is denoted by $\ell_i$ for each $i\in\Omega_x$.

\subsection{Problem Statement}
The goal of the paper is related to but different from the standard crowdsourcing problem which aims at learning a model that one can use to infer the true label $y_1,...,y_n$ using noisy annotations $L[\Omega]$ (and sometimes item features $X$).
Many highly practical models were proposed for that task already \citep{dawid1979maximum,welinder2010multidimensional,zhang2016spectral,zhou2015regularized,shah2020permutation}. 

Complementary to the existing work that mainly focuses on label inference,
we consider the problem of cost-saving. Specifically, we would like to design algorithms to reduce the expected number of new annotations needed to label a new item. The algorithm uses a pre-trained crowdsourcing model as well as the training dataset $X$ and $L[\Omega]$.

\subsection{Dawid-Skene Model and Score Functions}
The primary model that we work with in this paper is the Dawid-Skene (DS) model \citep{dawid1979maximum,zhang2016spectral}, which assumes the following data generating process.
\begin{enumerate}
    \item For each $j\in[n]$, $y_j\sim \rm{Categorical}(\tau).$
    \item For each $j\in[n],i\in[m]$, $\ell_{ij} \sim \rm{Categorical}(\mu_{\mathit{y_j,i}}).$
    \item We observe $\ell_{ij}$ with probability $\pi_{ij}$.
\end{enumerate}
where $\tau$ and $\mu_{y,i}$ denote the probability distributions defined on $[k]$. In particular, $\mu_{y,i}$ is the column $y$ of the \emph{confusion matrix} of worker $i$, which the DS model uses to describe $\P_i(\ell|y)$. We denote the confusion matrix associated with worker $i$ by $\mu_i \in\R^{k\times k}$.  
Once the DS model is learned, we can make use of the learned parameters $\tau$ and $\mu$ to infer the true labels using worker annotations via the posterior belief
\begin{equation}\label{eq:ds_inference}
P(y|\ell_1, \ell_2, ..., \ell_m) \propto P(y)\prod_{i=1}^{m}P(\ell_i | y)  =  \tau[y] \prod_{i=1}^m \mu_i[\ell_i,y].
\end{equation}
Take $\log$ for both sides and dropping the additive constant, we obtain the \emph{score function} that is induced by the DS model
\begin{equation}\label{eq:score_ds}
S_{\rm{DS}}(y|\ell_{i}\ \forall i \in \Omega_x )
= \log \tau[y] +\sum_{i=1}^{m}\log \mu_i[\ell_i,y].
\end{equation}
This is a weighted voting rule based on a pre-trained DS model. Similarly, we can cast the inference procedure of other crowdsourcing models as maximizing such a score function as well. For example, in the Majority Voting (MV) approach,
\begin{equation}\label{eq:score_mv}
    S_{\rm{MV}}(y|\Omega_x ) = \sum_{i=1}^m \mathbf{1}(\ell_{i}(x) = y),
\end{equation}
where $\mathbf{1}(\cdot)$ is the indicator function. Notably, no training datasets are needed for majority voting. The exposition above suggests that test time involves collecting a handful of worker annotations (choosing $\Omega_x$) and calculating a voting score specified by the crowdsourcing model in a form of 
\begin{equation}\label{eq:score_generic}
    S(y|\Omega_x ) = \sum_{i\in\Omega_x} S_i(y,\ell_i(x)).
\end{equation}
where $S_i$ is supplied by the model that connects annotation $\ell_i$ to label $y$. Then the label $y$ that maximizes the score is chosen.

\subsection{A Statistical Estimation Framework}
In the ideal world, when money is not a concern, we will poll \emph{all} workers and calculate
\begin{equation}\label{eq:IW_score}
S(y|[m]) = \sum_{i=1}^m S_i(y,\ell_i(x)).
\end{equation}
In practice, however, just as we cannot afford to poll all voters to estimate who is winning the presidential election, we cannot afford to poll everyone to annotate a single data point either. But do we have to?

Notice that we can frame the question as a classical point estimation problem in statistics, where the statistical quantity of interest is
\begin{equation}\label{eq:target}
v^x(y) :=  \E\bigg[ \frac{1}{m}\sum_{i=1}^m S_i(y,\ell_i(x))\bigg],
\end{equation}
the expectation of the ideal world score function \eqref{eq:IW_score}, rescaled by $1/m$. 
In the above, the expectation is taken over the randomness in worker's annotation. For example, if we select each worker independently with probability $\pi$, then the approach used in \eqref{eq:score_ds} and \eqref{eq:score_mv} would be an unbiased estimate of $v^x(y)$, if we rescale them by a factor of $\pi^{-1}$.

The advantage of translating the problem into a classical statistical estimation problem is that there is now a century of associated literature that we can tap into, including those on adaptive sampling and variance reduction techniques. We emphasize that while we will be using a crowdsourcing model, for example, the Dawid-Skene model, we do not assume that the data is generated according to the model. In fact, we are not imposing any restrictions on how workers annotate items, except that
\begin{enumerate}
	\item $\ell_{i}(x) \ \forall i\in[m]$ are mutually independent given any item $x$.
	\item $\Var[S_i(y,\ell_i(x))] < +\infty \quad \forall x,y$.
\end{enumerate}
These are very mild assumptions that are typically true in practice. It is generally difficult to analytically model human behaviors because it depends on how the item is presented to worker as well as the worker's knowledge and cognitive processes. The agnostic learning point of view helps disentangle the approximation-theoretic questions from the statistical question of estimating the best approximation possible using a given crowdsourcing model.

The remainder of the paper will be about designing estimators of $v^x(y)$ that achieves accurate label inference at a low cost and their corresponding theory and experiments. To avoid any confusions, we emphasize again that item $x$ is fixed. All the estimators are defined for each $y$ separately. $\ell_1,...,\ell_m$ are random variables that comes out of the unknown process of workers looking at the item $x$. Whenever the dependence is clear from context, we drop the conditioning on $x$ for better readability.

\section{Benchmark Approaches}\label{sec:bench}
In this section, we describe a few baseline approaches for estimating $v^x(y)$ and their corresponding cost in number of annotations.

\subsection{Ideal World (IW) Estimator}
In the ideal world, all workers are required to label $x$:
\begin{equation}\label{eq:IW}
\hat{v}_{\rm{IW}}(y)=\frac{1}{m}\sum_{i=1}^{m} S_i(y,\ell_i).
\end{equation}
This estimator incurs cost of $m$ and it is unbiased, with variance of $\frac{1}{m^2}\sum_{i=1}^m\Var[S_i(y,\ell_i(x))]$. This is arguably the best one can do with additional information.

\subsection{Importance Sampling (IS) Estimator}
A more affordable approach is to directly sample the workers. Specifically, we will include worker $i$ independently with probability $\pi_i$ \footnote{This is called a Poisson sampling \citep{sarndal2003model} in the survey sampling theory.}.
\begin{equation}\label{eq:IS}
\hat{v}_{\rm{IS}}(y) = \frac{1}{m}\sum_{i=1}^m \frac{\mathbf{1}(i\in \Omega)}{\pi_i}S_i(y,\ell_i).
\end{equation}
The expected cost of the IS estimator is $\sum_{i\in[m]}\pi_i$ and it is clearly an unbiased estimator.

\begin{thm}\label{thm:IS}
The $\hat{v}_{\rm{IS}}(y)$ is unbiased and
\begin{align*}
\Var[\hat{v}_{\rm{IS}}(y)] = \frac{1}{m^2}\bigg(\sum_{i=1}^m \frac{1}{\pi_i}\Var[S_i(y,\ell_i)]+(\frac{1}{\pi_i}-1)\E[S_i(y,\ell_i)]^2\bigg).
\end{align*}
\end{thm}
\begin{proof}
By the independence of sampling,
\begin{align*}
\E[\hat{v}_{\rm{IS}}(y)]&=\frac{1}{m}\sum_{i=1}^{m}\E \Big[\mathbf{1}(i\in \Omega) \frac{1}{\pi_i} S_i(y,\ell_i)\Big]\\
&=\frac{1}{m}\sum_{i=1}^{m}\pi_i \frac{1}{\pi_i} \E[S_i(y,\ell_i)]\\
&=v^x(y).
\end{align*}
To calculate the variance, we use the independence and then apply the law of total variance on each $i$:
\begin{align*}
\Var[\hat{v}_{\rm{IS}}(y)] =&\ \frac{1}{m^2}\sum_{i=1}^m \frac{1}{\pi_i^2}\Var[\mathbf{1}(i\in \Omega)S_i(y,\ell_i)]\\
=&\ \frac{1}{m^2}\sum_{i=1}^m \frac{1}{\pi_i^2} \Big(\E[(\mathbf{1}(i\in \Omega))^2] \E[(S_i(y,\ell_i))^2] - \E[\mathbf{1}(i \in \Omega)]^2 \E [S_i(y, \ell_i)]^2 \Big)\\
=&\ \frac{1}{m^2}\sum_{i=1}^m \frac{1}{\pi_i^2}\Big(\pi_i \Var[S_i(y,\ell_i)] +\pi_i \E[S_i(y,\ell_i)]^2 - \pi_i^2\E[S_i(y,\ell_i)]^2\Big)\\
=&\ \frac{1}{m^2}\bigg(\sum_{i=1}^m \frac{1}{\pi_i}\Var[S_i(y,\ell_i)] +(\frac{1}{\pi_i}-1)\E[S_i(y,\ell_i)]^2\bigg).
\end{align*}
\end{proof}
\begin{remark}If $\pi_i\equiv \pi$, then we are essentially doing the standard probabilistic inference as in \eqref{eq:score_ds} and \eqref{eq:score_mv}. When $\pi_i=1$, IS trivially subsumes IW \eqref{eq:IW} as a special case. Moreover, since $x$ is fixed, the sampling $\pi_i$ can be chosen as a function of the item $x$ without affecting the above results.
\end{remark}

\subsection{Direct Method (DM)}
Finally, there is an option that comes with no cost. Recall that we have a dataset $X$ and $L$ that were used to train the crowdsourcing model at our disposal. We can reuse the dataset and train $m$ supervised learners to imitate each worker's behavior. Let $\hat{\ell}_1,...,\hat{\ell}_m$ be the fictitious annotations provided by these supervised learners, we can simply plug them into the ideal world estimator \eqref{eq:IW} without any cost,
\begin{equation}\label{eq:DM}
\hat{v}_{\rm{DM}}(y)=\frac{1}{m}\sum_{i=1}^{m}\E[S_i(y,\hat{\ell}_i)].
\end{equation}
Following the convention in the contextual bandits literature \citep{jiang2016doubly}, we call this approach the direct method. The additional $\E$ is introduced to capture the case when supervised learner outputs a soft annotation $\hat{\ell}_i$.

The variance of this approach is $0$. However, as we mentioned previously, we can never hope to faithfully learn human behaviors, especially when we only have a small number of annotations in the training data for each worker $i$.  As a result, \eqref{eq:DM} may suffer from a bias that does not vanish even as $m\rightarrow \infty$.

\section{Main Results}\label{sec:drc}
In this section, we adapt an old statistical technique, doubly robust estimation, to crowdsourcing problem. 
\subsection{Doubly Robust Crowdsourcing}
As we established in the last section, IS estimator is unbiased but suffers from a large variance, especially when we would like to cut cost and use a small sampling probability.
The DM estimator incurs no additional annotation cost and has no variance, but it can potentially suffer from a large bias due to supervised learners not imitating the workers well enough. 

Doubly robust estimation \citep{rotnitzky1995semiparametric,dudik2014doubly} is a powerful technique that allows us to reduce the variance using a DM estimator while retaining the unbiasedness, hence getting the best of both worlds. The doubly robust estimator works as follows:
\begin{equation}\label{eq:DR}
\hat{v}_{\rm{DR}}(y)=\frac{1}{m}\sum_{i=1}^{m}\bigg(\E[S_i(y,\hat{\ell}_i)]+\frac{\mathbf{1}(i\in\Omega)}{\pi_i}( S_i(y,\ell_i)-\E[S_i(y,\hat{\ell}_i)])\bigg).
\end{equation}
The doubly robust estimator can be thought of using the DM as a baseline and then use IS to estimate and correct the bias. Provided that the supervised learners are able to provide a nontrivial approximation of the workers, the doubly robust estimator is expected to reduce the variance.
Just to give two explicit examples of $\hat{v}_{DR}(y)$, under the Dawid-Skene model, the doubly robust estimator is
\begin{equation}
\frac{1}{m}\sum_{i=1}^{m} \log P_{\mu_i}(\hat{\ell}_i | y) +\frac{\mathbf{1}(i\in \Omega)}{\pi_i} \log \frac{P_{\mu_i}(\ell_i | y)}{P_{\mu_i}(\hat{\ell}_i |y)}.
\end{equation}
Similarly, for the majority voting model, we can write
\begin{align}
\frac{1}{m} \sum_{i=1}^m e_{\hat{\ell}_i}+\frac{1}{\pi_i} \mathbf{1}(i\in \Omega) (e_{\ell_i} - e_{\hat{\ell}_i}),
\end{align}
where $e_\ell$ is the basis vector where $\ell$ indicates the location of $1$, otherwise $0$.

\begin{thm}[DRC] \label{cor:drc} The doubly robust estimator \eqref{eq:DR} is unbiased and its variance is:
\begin{align*}
    \frac{1}{m^2}\bigg(\sum_{i=1}^m \frac{1}{\pi_i}\Var[S_i(y,\ell_i)] +\Big(\frac{1}{\pi_i}-1 \Big)\E[S_i(y,\ell_i)-S_i(y,\hat{\ell}_i)]^2\bigg).
\end{align*}
\end{thm}
\begin{proof-sketch}
Note that the first part  $\frac{1}{m}\sum_{i=1}^{m}\E[S_i(y,\hat{\ell}_i)]$
of the estimator is not random. The result follows directly by invoking Theorem~\ref{thm:IS} on the second part of the estimator, which is an importance sampling estimator of the bias. 
\end{proof-sketch}
\begin{proof}
By the independence of sampling,
\begin{align*}
\E[\hat{v}_{\rm{DR}}(y)]&=\frac{1}{m}\sum_{i=1}^{m} \E \big[\E [S_i(y, \hat{\ell}_i)]\big] + \frac{1}{\pi_i} \E \big[\mathbf{1}(i\in \Omega) (S_i(y,\ell_i) - \E[S_i (y, \hat{\ell}_i)])\big]\\
&=\frac{1}{m}\sum_{i=1}^{m} \E \big[\E [S_i(y, \hat{\ell}_i)]\big] + \pi_i \frac{1}{\pi_i} \E[S_i(y,\ell_i) - \E[S_i (y, \hat{\ell}_i)]]\\
&= \frac{1}{m}\sum_{i=1}^{m} \E[S_i(y,\ell_i)]\\
&=v^x(y).
\end{align*}
To calculate the variance, we use the independence and then apply the law of total variance on each $i$:
\begin{align*}
\Var[\hat{v}_{\rm{DR}}(y)] =&\ \frac{1}{m^2}\sum_{i=1}^m \frac{1}{\pi_i^2}\Var \big[\mathbf{1}(i\in \Omega) (S_i(y,\ell_i) -\E[S_i (y, \hat{\ell}_i)])\big]\\
=&\ \frac{1}{m^2}\sum_{i=1}^m \frac{1}{\pi_i^2} \Big(\E[(\mathbf{1}(i\in \Omega))^2] \E[(S_i(y,\ell_i) -\E[S_i (y, \hat{\ell}_i)])^2]\\
&\quad - \E[\mathbf{1}(i \in \Omega)]^2 \E [S_i(y, \ell_i) -\E[S_i (y, \hat{\ell}_i)]]^2 \Big)\\
=&\ \frac{1}{m^2}\sum_{i=1}^m \frac{1}{\pi_i^2}\Big(\pi_i \Var[S_i(y,\ell_i)] +\pi_i \E[S_i(y,\ell_i)-\E[S_i (y, \hat{\ell}_i)]]^2\\
&\quad - \pi_i^2\E[S_i(y,\ell_i)-\E[S_i (y, \hat{\ell}_i)]]^2\Big)\\
=&\ \frac{1}{m^2}\bigg(\sum_{i=1}^m \frac{1}{\pi_i}\Var[S_i(y,\ell_i)] +\Big(\frac{1}{\pi_i}-1\Big)\E[S_i(y,\ell_i)-S_i (y, \hat{\ell}_i)]^2\bigg).
\end{align*}
\end{proof}
\begin{remark}
First, if workers are deterministic, the first part of the variance $\Var[S_i(y,\ell_i)]\equiv 0$. Second, if the supervised learner imitates workers perfectly \emph{in expectation}, the second part of the variance vanishes. Finally and most importantly, the supervised learner does \emph{not} have to be perfect. In the simple case of a deterministic workers, the percentage of agreements between supervised learners and their human counterparts directly translate into a reduction of the variance of about the same percentage, for free.
\end{remark}
The third point is especially remarkable as it implies that even a trivial surrogate that outputs a label at random could lead to a $1/k$ factor reduction of the variance. In addition, a good set of worker imitators with $90\%$ accuracy can lead to an order of magnitude smaller variance and hence allow us to incur a much lower cost on average. We will illustrate the effects of doubly robust estimation more extensively in the experiments. This feature ensures that our proposed method remains applicable even in the case when the training dataset contain few annotations from some subset of the features.

\subsection{Confidence-Based Adaptive Sampling}\label{subsec:apt}
Doubly robust estimation allows us to reduce the variance. However, doubly robust is still an importance sampling-based method that requires the number of new annotations to be at least linear in the number of data points to label. In this section, we propose using supervised worker imitation to obtain confidence estimates for free and using them to construct confidence-based adaptive sampling schemes.

We propose two rules.
\begin{enumerate}
    \item \textbf{Adaptive item selection} For each new data point, run DM first. If DM predicts label $y$ with an overwhelming confidence, then chances are, there is no need to collect more annotations. If not, human workers are needed.
    \item \textbf{Adaptive worker selection} We can adaptively choose which worker to annotate a given item. Instead of sampling at random with probability $\pi$, we choose a set of adaptive sampling probability $\pi_1, \cdots, \pi_m$ that makes high confidence workers more likely to be selected. As different workers have different skill sets, confidence may depend strongly on each item $x$. We propose to calculate such item-dependent confidence using outcome of the imitated workers and the confusion matrices from the DS model.
\end{enumerate}

In both cases, we need a way to measure confidence given a probability distribution. A threshold is introduced to decide whether accept predicted labels or not \citep{branson2017lean}. Margin in multi-class classification is defined as the difference between the score of true label and the largest score of other labels \citep{mohri2012foundations}. Inspired by them, we define confidence margin of a probability as follows.
\begin{defn}[Confidence Margin]
Given a discrete probability distribution $\pi_1, ..., \pi_m$, its confidence margin $\rho$ is defined as the difference between the largest probability and the second largest one.
\end{defn}
Based on confidence margin, we propose three new methods: DRC with Adaptive Item Selection (DRC-AIS), DRC with Adaptive Worker Selection (DRC-AWS), and the combination DRC-AWS-AIS. In DRC-AIS, DM is performed on all labels. For each item, the surrogate label given by DM follows \eqref{eq:ds_inference} to get the posterior belief, which describes how confidently DM gives the label of this item. Based on posterior belief, its confidence margin $\rho_\text{AIS}$ is compared with the given confidence margin parameter $\rho$. If $\rho_\text{AIS}$ is larger, DRC-AIS takes the surrogate label provided by DM with no worker cost, otherwise, DRC-AIS follows the regular DRC model which incures cost. In DRC-AWS, again DM runs first and gets surrogate labels $\hat{\ell}_i$. Then from each worker $i$'s confusion matrix we can get the labeling probability $P(\hat{\ell}_i|y)$, whose confidence margin is used as the worker score $\gamma_i$ and $\gamma_1, \cdots, \gamma_m$ are normalized to be a distribution. For each item, worker $i$ will be sampled with probability $\gamma_i$. However, in this case the sampling probability for each worker is usually very small, thus, we can introduce a parameter $\lambda$ to multiply with $\gamma_i$ to increase the sampling probability.
If a worker is sampled, the corresponding label is used as regular DRC model.

Table \ref{tab_comp} shows the summary of benchmark and DRC approaches.
As we can see, IW and IS take the fewest input elements while IW uses the most worker cost. DM uses no cost because it only takes advantage of surrogate labels given by worker imitation. DRC has similar cost as IS while it is expected to improve the ground truth inference with less variance. Thanks to the confidence-based adaptive sampling techniques, DRC-AIS and DRC-AWS are able to save more cost than DRC.

\begin{table}[!htbp]
	\centering
	\caption{Summary of benchmark and DRC approaches in DS models. $L, X$ are annotation matrix and item feature matrix. $\mu, \tau$ are probability distributions from DS models. $f$ denotes the supervised classifier for worker imitation.}
	\label{tab_comp}
	\begin{tabular}{cccc}
		\noalign{\smallskip} \hline \noalign{\smallskip}
		\textbf{Methods}                  & \textbf{Input} & \textbf{Sampling}           & \textbf{Cost}                    \\ \noalign{\smallskip} \hline \noalign{\smallskip}
		IW         & $L$              & No                 & $O(nm)$                   \\ \noalign{\smallskip} \hline \noalign{\smallskip}
		IS & $L$              & $\pi$ & $O(\pi nm)$ \\ \noalign{\smallskip} \hline \noalign{\smallskip}
		DM              & $X+f$         & No & $0$                       \\ \noalign{\smallskip} \hline \noalign{\smallskip}
		DRC       & $L+\mu+\tau(y)+f$       & $\pi$ & $O(\pi nm)$ \\ \noalign{\smallskip} \hline \noalign{\smallskip}
		DRC-AIS       & $L+\mu+\tau(y)+f$       & $\pi$ & $O(\pi n^*m)$ \\ \noalign{\smallskip} \hline \noalign{\smallskip}
		DRC-AWS       & $L+\mu+\tau(y)+f$       & $\pi_{1:m}$& $O(\sum_{i\in[m]}\pi_i n)$ \\ \noalign{\smallskip} \hline \noalign{\smallskip}
	\end{tabular}\\
	\scriptsize{$n^* \ll n$ denotes number of items AIS selected as low-confidence.}
\end{table}

\subsection{Weight-Clipping in DRC}
Adaptive worker selection involves making selection probability $\pi$ larger for some workers while smaller for others. According to Theorem~\ref{cor:drc}, the variance is proportional to $\sum_i\pi_i^{-1}$, hence even a single $\pi_i$ being close to $0$ would result in a huge variance. In off-policy evaluation \citep{wang2017optimal} problems this issue is addressed by clipping the importance weight at a fixed threshold $\eta$. This results in the clipped doubly robust estimator.
\begin{align}\label{eq:DR_clipped}
&\hat{v}_{\rm{DR}_{\eta}}(y) =\frac{1}{m}\sum_{i=1}^{m}\Big(\E[S_i(y,\hat{\ell}_i)]+ \mathbf{1}(i\in\Omega) \min\{\eta, \pi_i^{-1}\}( S_i(y,\ell_i)-\E[S_i(y,\hat{\ell}_i)])\Big).
\end{align}
Its bias and variance are given as follows.
\begin{thm}
The clipped doubly robust estimator obeys that:
\begin{align*}
\mathrm{Bias}(\hat{v}_{\rm{DR}_\eta}(y)) &= \bigg|\frac{1}{m}\sum_{i=1}^m
\min\{\pi_i\eta-1,0\}\E[S_i(y,\ell_i)-S_i(y,\hat{\ell}_i)] \bigg|.\\
\Var[\hat{v}_{\rm{DR}_\eta}(y)] &= \frac{1}{m^2}\sum_{i=1}^m \min\{\eta^2\pi_i^2,1\}\bigg( \frac{1}{\pi_i}\Var[S_i(y,\ell_i)] +\Big(\frac{1}{\pi_i}-1\Big)\E[S_i(y,\ell_i)-S_i(y,\hat{\ell}_i)]^2\bigg).
\end{align*}
\end{thm}
\begin{proof}
By definition of bias,
\begin{align*}
\mathrm{Bias} (\hat{v}_{\rm{DR}_\eta}(y)) =&\ \Big|\E[\hat{v}_{\mathrm{DR}_\eta} (y)] - \frac{1}{m}\sum_{i=1}^m \E[S_i(y, \ell_i)]\Big|\\
=&\ \bigg| \frac{1}{m}\sum_{i=1}^m \Big(\E[\E[ S_i(y, \hat{\ell}_1)]] + \E \big[\mathbf{1}(i \in \Omega) \min\{\eta, \pi_i^{-1}\} (S_i(y,\ell_i)- \E[S_i (y, \hat{\ell}_i)])\big]\\
&\quad - \E[S_i(y,\ell_i)]\Big)\bigg|\\
=&\ \bigg| \frac{1}{m}\sum_{i=1}^m \Big(\E[ S_i(y, \hat{\ell}_1)] + \min\{\pi_i \eta, 1\} \E [S_i(y,\ell_i)- \E[S_i (y, \hat{\ell}_i)]]- \E[S_i(y,\ell_i)]\Big)\bigg|\\
=&\ \bigg|\frac{1}{m}\sum_{i=1}^m
\min\{\pi_i\eta-1,0\}\E[S_i(y,\ell_i)-S_i(y,\hat{\ell}_i)] \bigg|.
\end{align*}
And the variance can be calculated as,
\begin{align*}
\Var[\hat{v}_{\rm{DR}_\eta}(y)] =&\ \frac{1}{m^2}\sum_{i=1}^m \min \{\eta^2,\pi_i^{-2}\}\Var [\mathbf{1}(i \in \Omega) (S_i(y, \ell_i) - \E[S_i(y, \hat{\ell}_i)])]\\
=&\ \frac{1}{m^2}\sum_{i=1}^m \min \{\eta^2,\pi_i^{-2}\} \Big(\E[(\mathbf{1}(i\in \Omega))^2] \E[(S_i(y,\ell_i) -\E[S_i (y, \hat{\ell}_i)])^2]\\
&\quad - \E[\mathbf{1}(i \in \Omega)]^2 \E [S_i(y, \ell_i) -\E[S_i (y, \hat{\ell}_i)]]^2 \Big)\\
=&\ \frac{1}{m^2}\sum_{i=1}^m \min \{\eta^2,\pi_i^{-2}\} \Big(\pi_i \Var[S_i(y,\ell_i)] +\pi_i \E[S_i(y,\ell_i)-\E[S_i (y, \hat{\ell}_i)]]^2\\
&\quad - \pi_i^2\E[S_i(y,\ell_i)-\E[S_i (y, \hat{\ell}_i)]]^2\Big)\\
=&\ \frac{1}{m^2}\sum_{i=1}^m \min\{\eta^2\pi_i^2,1\}\bigg( \frac{1}{\pi_i}\Var[S_i(y,\ell_i)] +(\frac{1}{\pi_i}-1)\E[S_i(y,\ell_i)-S_i(y,\hat{\ell}_i)]^2\bigg).
\end{align*}
\end{proof}
\begin{remark}
The bias bound indicates that only those workers we clipped who contribute to the bias. The variance bound implies that the part of variance from Worker $i$ is reduced to $O(\min\{1/\pi_i, \eta^2\pi_i\})$ from $O(1/\pi_i)$.
If the total amount of additional $\mathrm{Bias}^2$ introduced by the clipping is smaller than the corresponding savings in the variance, then clipping makes the estimator more accurate in Mean Squared Error (MSE). The theory inspires us to design an algorithm to automatically choose the threshold.
\end{remark}
\begin{remark}[Automatic choice of threshold]
Assume $\ell_{1},...,\ell_m$ are deterministic, $|\E[S_i(y,\ell_i)-S_i(y,\hat{\ell}_i)]| \leq \epsilon$. Then the bias of $\hat{v}_{\rm{DR}_\eta}(y)$ can be bounded by $\epsilon\sum_{i}\mathbf{1}(\pi_i^{-1}>\eta)/m$, and variance of $\hat{v}_{\rm{DR}_\eta}(y)$ can be bounded by $\epsilon^2\eta^2/m$.
Recall that MSE can be decomposed in to $\mathrm{Bias}^2 + \Var{}$. 
The optimal choice of $\eta$ that minimizes this upper bound is the one that minimizes 
$|\eta - \sum_{i}\mathbf{1}(\pi_i^{-1}>\eta) /\sqrt{m}|$.
This can be found numerically in time $O(m\log(m))$ by sorting $[\pi_1^{-1},...,\pi_m^{-1}]$ and applying binary search.
\end{remark}

\section{Experiments}\label{sec:exp}
In this section, we report our experimental results, including both synthetic and real-world experiments.

\subsection{Synthetic Experiments}
First we describe our experimental settings and then move to synthetic experiment results.

\subsubsection{Experimental Settings}
\begin{figure}[!htbp]
	\centering
	\includegraphics[width=5in]{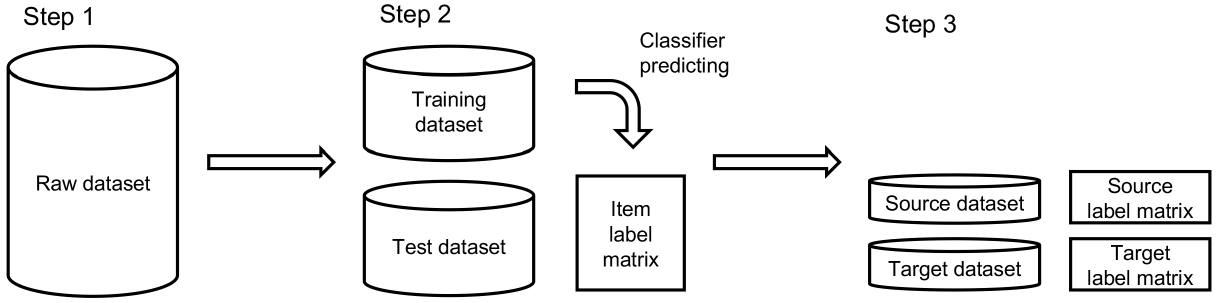}
	\caption{Workflow of generating crowdsourcing datasets with item features.\label{fig-exp}}
\end{figure}

We are using supervised classification datasets to do synthetic experiments. Due to absence of labeling matrix, we follow the workflow shown in Figure \ref{fig-exp} to do worker imitation to generate crowdsourcing datasets, which has three steps.
\begin{enumerate}
    \item We starts with the raw dataset in step 1. If the dataset was split into training and test parts, we combine them together.
    \item In step 2, we uniformly sample the dataset into two equal parts, one for training and the other for test. In order to remove the randomness of this splitting process, the sampling index is fixed and saved for all further experiments. Training means we use this part of data to train $m$ decision trees to simulate the generating process of crowdsourcing labels. For one item only $\sqrt{d}$ features, a subset of all $d$ features, can be observed by each tree where $d$ is the total number of features. Then these $m$ decision trees are used to make predictions on the test set to obtain the item label matrix.
    \item In step 3, we uniformly sample the test part of step 2 into two equal parts, one working as source part and the other working as target part. For all experiments, this sampling process will be repeated $20$ times.
    Also, $m$ classifiers, one for each worker, are trained on source dataset to simulate worker behaviors to give surrogate labels for all DRC approaches. For all synthetic experiments, $m=50$. Evaluations are performed on the target label matrix.
\end{enumerate}

We use five classification datasets, Segment, Satimage, Usps \citep{hull1994a}, Pendigits, and Mnist \citep{lecun1998gradient}, collected by Libsvm \citep{chang2011libsvm}, which are all publicly available. Table \ref{tab:dataset_syn} shows their statistics of test dataset and item label matrix of Step 2.

\begin{table}[!htbp]
\centering
	\caption{Statistics of synthetic datasets.}\label{tab:dataset_syn}
	\begin{tabular}{ccccc}
		\noalign{\smallskip} \hline \noalign{\smallskip}
\textbf{Dataset} & \textbf{\# item} & \textbf{\# worker} & \textbf{\# dimension} & \textbf{\# class} \\ \noalign{\smallskip} \hline \noalign{\smallskip}
Segment  &  $1,155$            &  $50$         & $19$          &  $7$        \\ \noalign{\smallskip} \hline \noalign{\smallskip}
Satimage     & $3,217$             & $50$          &  $36$            &  $6$        \\ \noalign{\smallskip} \hline \noalign{\smallskip}
Usps  & $4,649$        & $50$      & $256$        & $10$   \\ \noalign{\smallskip} \hline \noalign{\smallskip}
Pendigits  & $5,496$        & $50$      & $16$        & $10$   \\ \noalign{\smallskip} \hline \noalign{\smallskip}
Mnist  & $35,000$        & $50$      & $780$        & $10$   \\ \noalign{\smallskip} \hline \noalign{\smallskip}
\end{tabular}
\end{table}

All experimental results, in figures or tables, are presented after repeating $20$ times with $98\%$ asymptotic confidence interval of the expected accuracy based on inverting Wald's test, that is,
\begin{equation}
    \mu \pm 1.96 \sigma /\sqrt{20},
\end{equation}
where $\mu, \sigma$ are the mean and stand error of accuracy. Based on Wald's test, statistical conclusions can be made with $98\%$ confidence.

\subsubsection{Algorithm Comparison}
In order to show our approach DRC is able to infer true labels with low worker cost, we compare our DRC approach with Importance Sampling (IS). In particular, we do experiments with both Dawid-Skene (DS) model and Majority Voting (MV) model. Therefore, we are comparing DRC-DS, DS, DRC-MV, and MV. Moreover, we have Direct Method (DM) and Classifiers trained on Inferred Labels (CI) as baselines. CI means classifiers are trained with inferred labels from source part and then make predictions for the target part, thus incurring no labeling cost. In detail, we do Poisson sampling over workers with $\pi$ going from $0.1$ to $1.0$ with interval of $0.1$. Decision trees with maximum depth $3$ are used to generate surrogate labels for DRC-DS, DRC-MV, and DM.

\begin{figure}[t]
	\centering
	\begin{minipage}{0.32\linewidth}\centering
		\includegraphics[width=\textwidth]{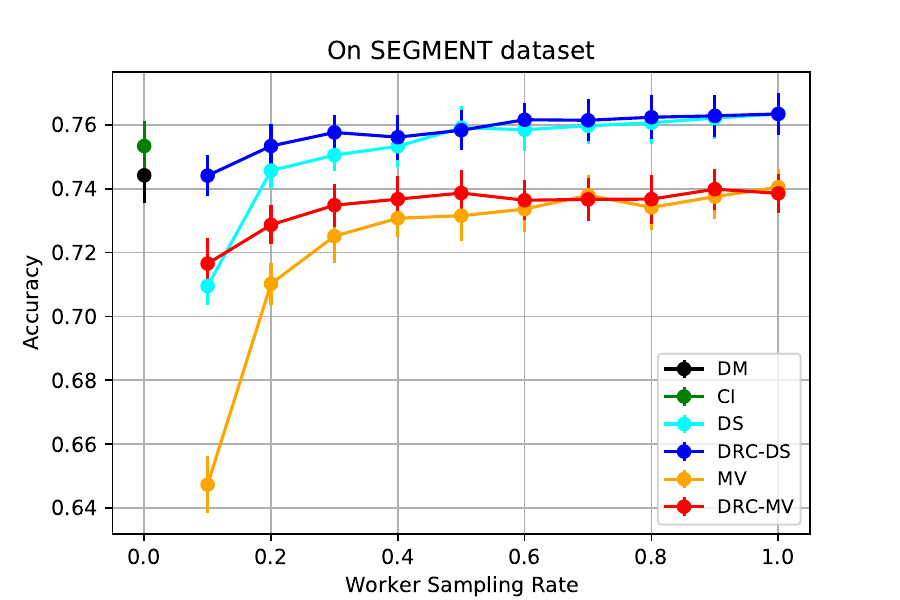}
	\end{minipage} 
	\begin{minipage}{0.32\linewidth}\centering
		\includegraphics[width=\textwidth]{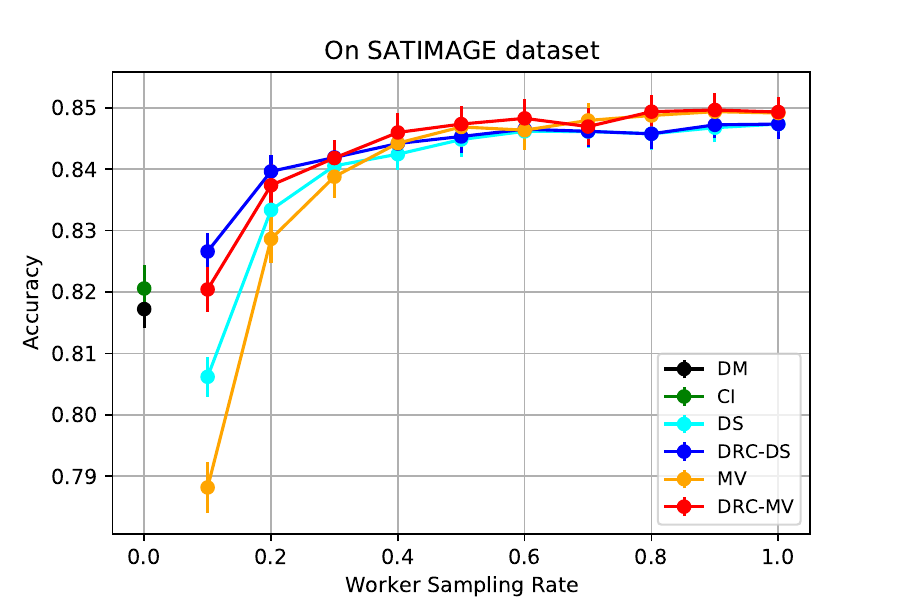}
	\end{minipage}
	\begin{minipage}{0.32\linewidth}\centering
		\includegraphics[width=\textwidth]{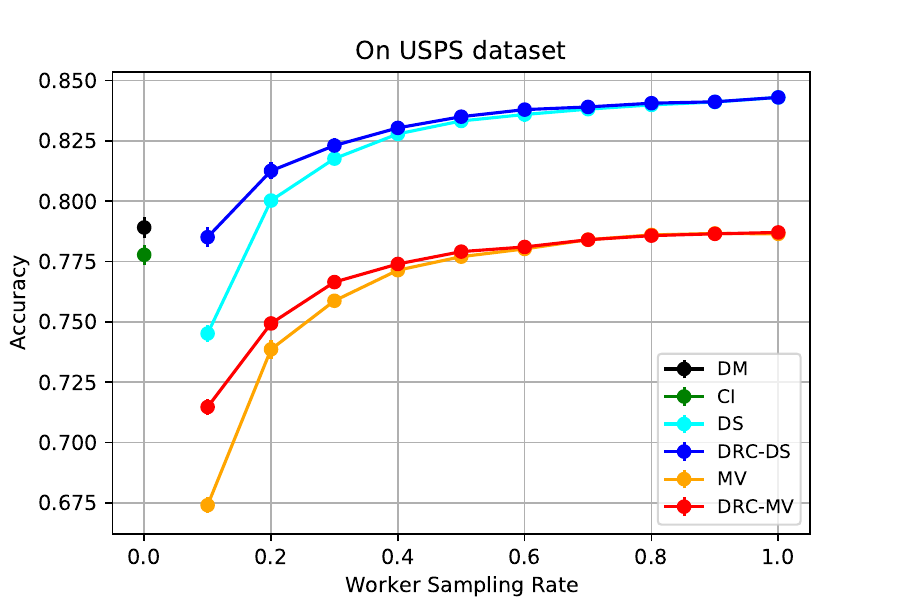}
	\end{minipage}
	\begin{minipage}{0.32\linewidth}\centering
		\includegraphics[width=\textwidth]{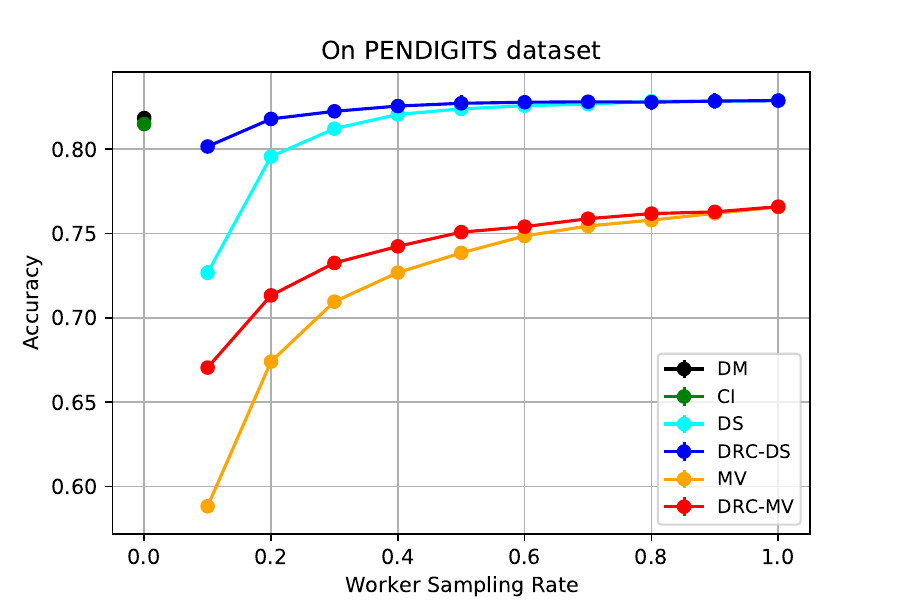}
	\end{minipage}
	\begin{minipage}{0.32\linewidth}\centering
		\includegraphics[width=\textwidth]{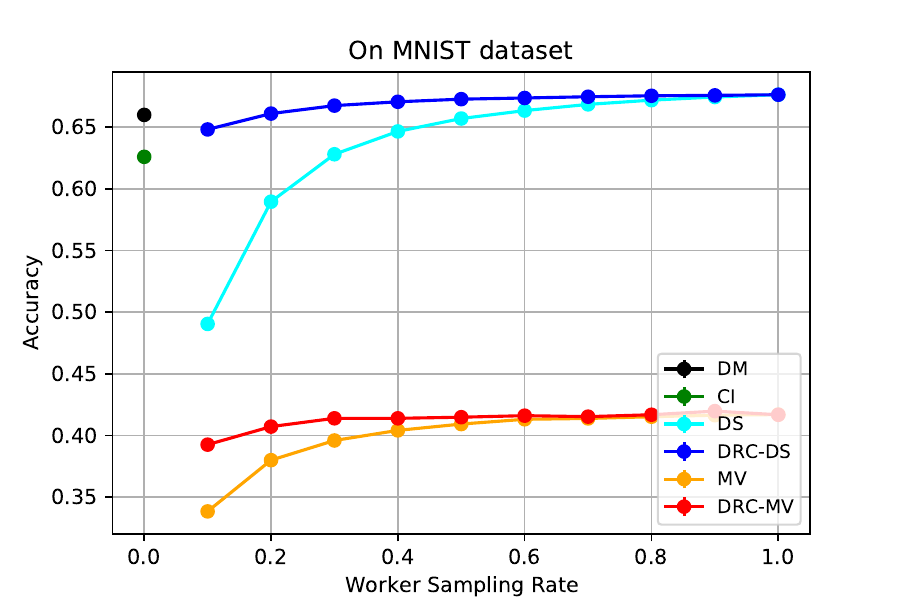}
	\end{minipage}
	\caption{Performances of DRC-DS, DRC-MV, DS, MV, and DM.\label{fig-exp1-cost}}
\end{figure}

Results are shown in Figure \ref{fig-exp1-cost}, which have three observations.
\begin{itemize}
    \item Given the same worker sampling rate, DRC-DS outperforms DS and DRC-MV works better than MV on all datasets, which shows the effectiveness of DRC and matches our theoretical understanding. Moreover, our DRC approaches work well with very few workers and then perform better with sampling rate increasing.
    \item Because CI and DM involve no worker cost, they are two nodes in the figures corresponding to $\pi=0$. Due to high bias incurred, DM performs unstably across all datasets. It performs well in some cases, but poorly on Satimage and Usps.
    \item On all datasets, except Satimage, DS performs better than MV. However, performance comparison between DS and MV is out of scope of this paper. We are focusing on improving existing approaches with DRC method.
\end{itemize}

\subsubsection{Effectiveness of AIS and AWS}
To show effectiveness of AIS and AWS, we compare four methods: DRC-DS, DRC-AIS, DRC-AWS, and DRC-AWS-AIS, while DM is used as the baseline. AWS and AIS rules are expected to save a lot of worker cost, so logarithmic worker cost is used, where cost is defined as number of workers that per item use. For DRC-AIS and DRC-AWS-AIS, the confidence margin parameter $\rho$ is set to be $0.03$, $0.06$, and $0.09$. For DRC-AWS and DRC-AWS-AIS, the multiplier parameter $\lambda$ is set to be $1$, $2$, $3$, $4$, $5$, $7$, $10$, $15$, $25$, and $50$.

\begin{figure}[!htbp]
	\centering
	\begin{minipage}{0.32\linewidth}\centering
		\includegraphics[width=\textwidth]{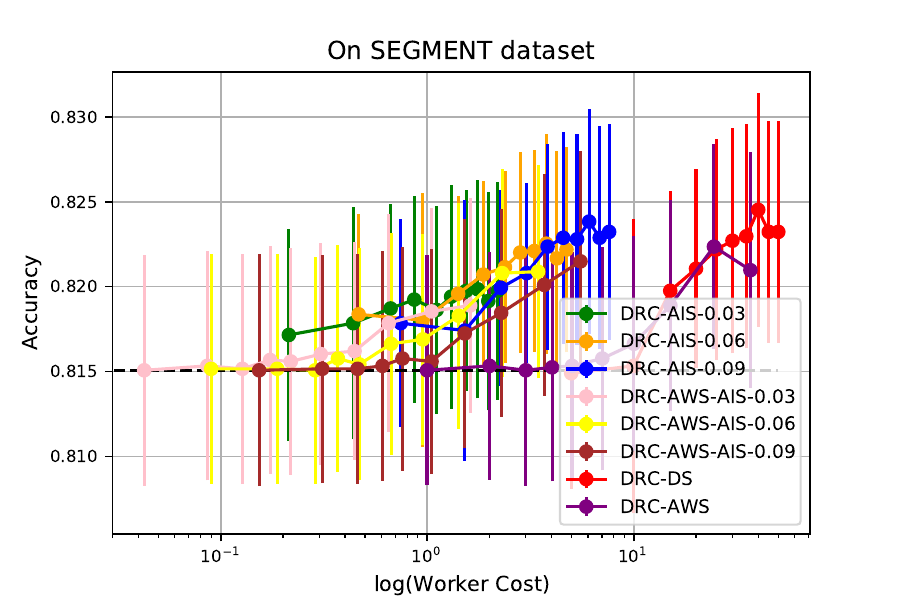}
	\end{minipage}
	\begin{minipage}{0.32\linewidth}\centering
		\includegraphics[width=\textwidth]{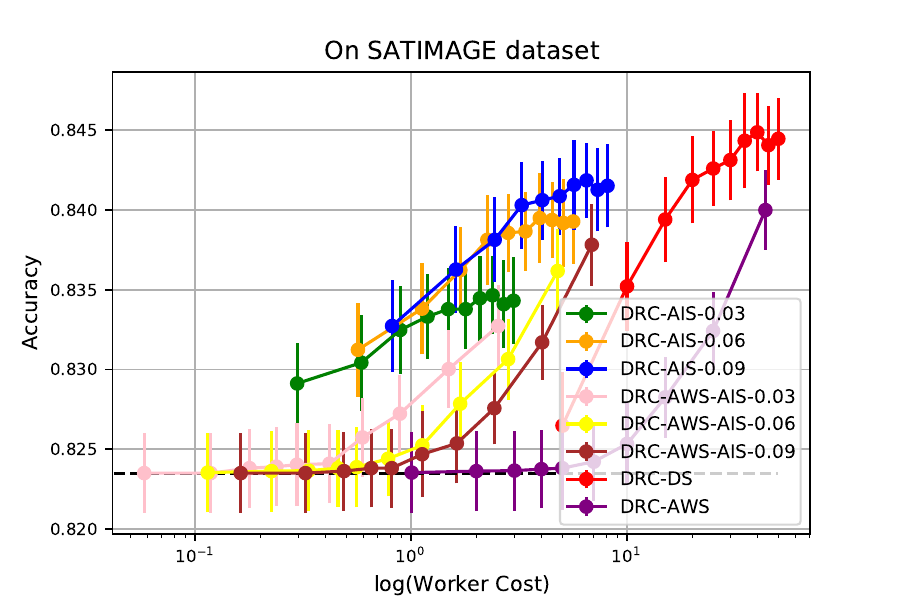}
	\end{minipage}
	\begin{minipage}{0.32\linewidth}\centering
		\includegraphics[width=\textwidth]{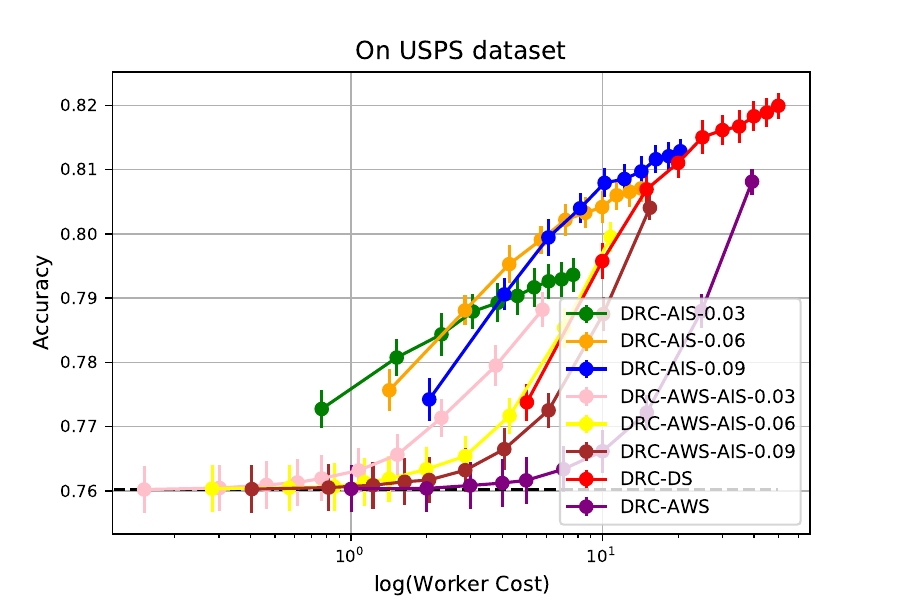}
	\end{minipage}
	\begin{minipage}{0.32\linewidth}\centering
		\includegraphics[width=\textwidth]{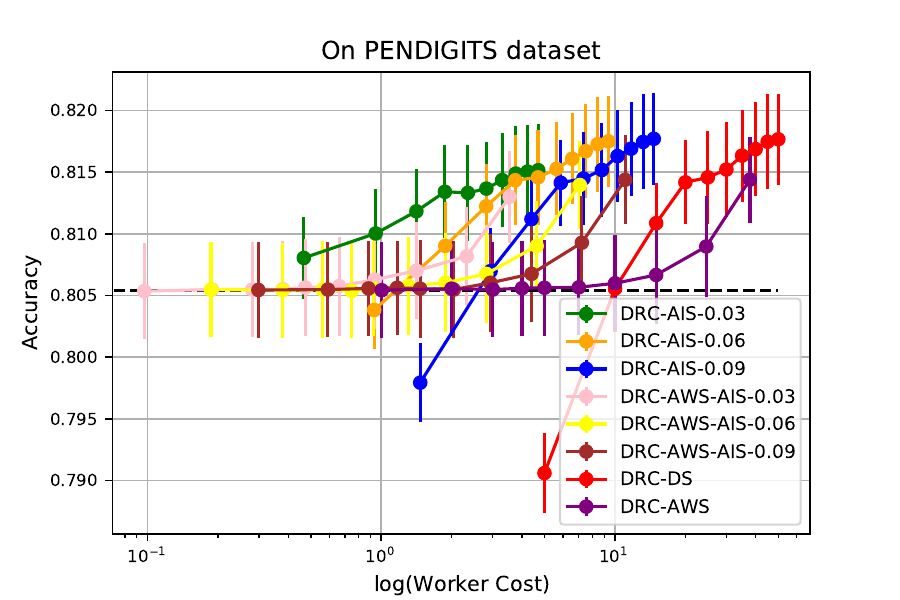}
	\end{minipage}
	\begin{minipage}{0.32\linewidth}\centering
		\includegraphics[width=\textwidth]{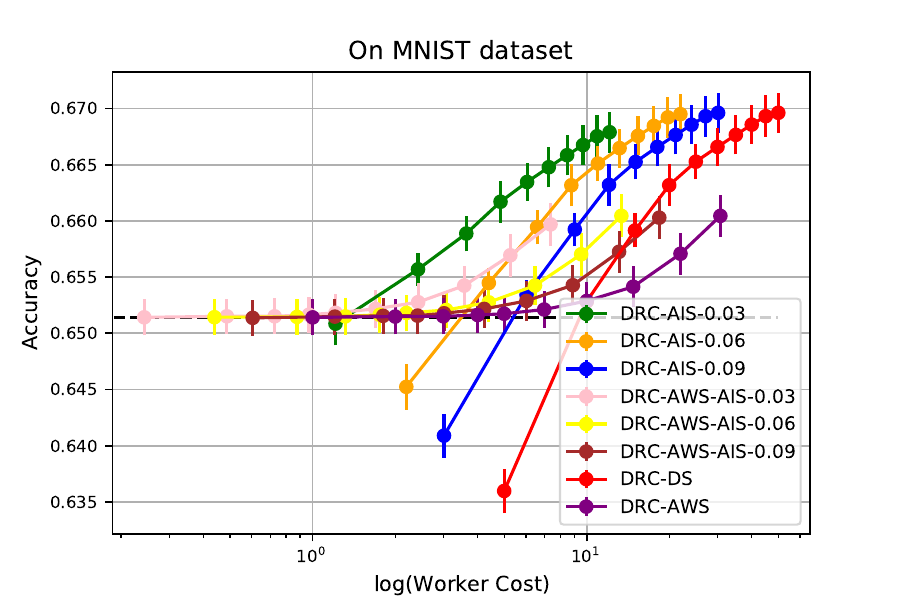}
	\end{minipage}
	\caption{Performances of DRC-DS, DRC-AIS, DRC-AWS, and DRC-AWS-AIS.\label{fig-exp2-cost}}
\end{figure}

With dashed black lines being performances of DM, results are shown in Figure \ref{fig-exp2-cost}, which have five observations:
\begin{itemize}
    \item Compared with DRC-DS, all AIS/AWS approaches are able to save labeling cost while maintaining almost the same accuracy, which validates AIS and AWS do play key roles in improving inference accuracy and saving worker cost at the same time.
    \item Among all four methods, DRC-AWS-AIS enjoys the lowest worker cost, which means AIS and AWS can work together.
    \item There is an accuracy-cost tradeoff for all approaches. All performances can be improved by introducing more worker cost, that is, greater $\rho, \lambda$.
    \item DRC-AWS and DRC-DS enjoy the same performance pattern while DRC-AWS-AIS and DRC-AIS have the same one. Specifically, because DM is applied at the first stage of algorithms, performances of DRC-AIS and DRC-AWS-AIS increase from the DM when worker cost is very small.
    \item No statistical conclusions can be made on Segment dataset due to high error bar.
\end{itemize}

\subsubsection{Ablation Study on Model Misspecification}
As mentioned in experimental settings, decision trees are used to generate the crowdsourcing datasets with item features, however, in real-world tasks, we have no idea of the label generating process. Therefore, model misspecification must be studied. In detail, Decision Trees (DT), Logistic Regression (LR), and Gaussian Naive Bayes (GNB) are used as supervised classifiers for DRC-DS, DRC-MV, and DM. $\pi=0.5$ is set over source label matrix. All parameters of DT, LR, and GNB are set as default according to \texttt{sklearn} package. Other experimental settings are the same as before.

In Table \ref{app-tab-r-unr}, based on Wald's test, statistically better results are set in \textbf{bold} fonts. Decision trees perform only slightly better than other two classifiers, which is good because it show our approaches are able to work without knowing the label generating process. In real crowdsourcing problem, it is of a high chance that we have no idea of this process.

\begin{table}[!htbp]
	\centering
	\caption{Performances of DRC-DS, DRC-MV, and DM with different supervised classifiers.}
	\label{app-tab-r-unr}
	\begin{tabular}{ccccc}
		\noalign{\smallskip} \hline \noalign{\smallskip}
		\textbf{Dataset}                                  & \textbf{Method} & \textbf{DT} & \textbf{LR} & \textbf{GNB} \\ \noalign{\smallskip} \hline \noalign{\smallskip}
		\multirow{3}{*}{Segment}                 & DRC-DS     & $0.7385\pm0.0065$   & $0.7425\pm0.0063$         & $0.7386\pm0.0066$          \\
		& DRC-MV    & $0.7731\pm0.0053$   & $0.7698\pm0.0064$         & $0.7677\pm0.0053$         \\
		& DM    & $0.7419\pm0.0061$   & $\mathbf{0.7788}\pm0.0084$         & $0.6969\pm0.0084$          \\ \noalign{\smallskip} \hline \noalign{\smallskip}
		\multirow{3}{*}{Satimage}                & DRC-DS     & $0.8483\pm0.0021$   & $0.8452\pm0.0028$        & $0.8421\pm0.0034$          \\
		& DRC-MV    & $\mathbf{0.8470}\pm0.0025$   & $0.8416\pm0.0029$         & $0.8415\pm0.0030$          \\
		& DM    & $0.8478\pm0.0016$   & $0.8270\pm0.0024$         & $0.7940\pm0.0033$          \\\noalign{\smallskip} \hline \noalign{\smallskip}
		\multirow{3}{*}{Usps}                    & DRC-DS     & $0.8323\pm0.0024$   & $0.8293\pm0.0020$         & $0.8260\pm0.0017$          \\
		& DRC-MV    & $0.8067\pm0.0023$   & $0.8022\pm0.0029$         & $0.7963\pm0.0018$          \\
		& DM    & $0.8315\pm0.0021$   & $\mathbf{0.8374}\pm0.0022$         & $0.7740\pm0.0034$          \\\noalign{\smallskip} \hline \noalign{\smallskip}
		\multirow{3}{*}{Pendigits}                    & DRC-DS     & $\mathbf{0.8196}\pm0.0021$   & $0.8156\pm0.0018$         & $0.8156\pm0.0020$          \\
		& DRC-MV    & $\mathbf{0.7359}\pm0.0018$   & $0.7274\pm0.0027$         & $0.7266\pm0.0023$          \\
		& DM    & $0.8193\pm0.0020$   & $\mathbf{0.8259}\pm0.0024$       & $0.8223\pm0.0017$          \\\noalign{\smallskip} \hline \noalign{\smallskip}
		\multirow{3}{*}{Mnist}                    & DRC-DS     & $0.6470\pm0.0020$   & $0.6426\pm0.0022$         & $0.5939\pm0.0017$          \\
		& DRC-MV    & $\mathbf{0.5317}\pm0.0013$   & $0.5240\pm0.0015$         & $0.4997\pm0.0011$          \\
		& DM    & $0.6471\pm0.0019$   & $0.6474\pm0.0021$         & $0.4240\pm0.0035$          \\\noalign{\smallskip} \hline \noalign{\smallskip}              
	\end{tabular}
\end{table}

\begin{table}[!htbp]
\centering
	\caption{Statistics of real-world datasets.}\label{tab:dataset_real}
	\begin{tabular}{ccccc}
		\noalign{\smallskip} \hline \noalign{\smallskip}
\textbf{Dataset} & \textbf{\# item} & \textbf{\# worker} & \textbf{\# dimension} & \textbf{\# class} \\ \noalign{\smallskip} \hline \noalign{\smallskip}
Music   &  $700$            &  $44$         & $124$          &  $10$        \\ \noalign{\smallskip} \hline \noalign{\smallskip}
Dog     & $798$             & $109$          &  $5,376$            &  $4$        \\ \noalign{\smallskip} \hline \noalign{\smallskip}
Tomato  & $4,999$        & $203$      & $1,200$        & $2$   \\ \noalign{\smallskip} \hline \noalign{\smallskip}
\end{tabular}
\end{table}

\subsection{Real-world Experiments}
Following synthetic experiments, we do algorithm comparison and study the effectiveness of AIS and AWS on real-world experiments.
\subsubsection{Algorithm Comparison}
We do experiments on three real-world datasets: Music Genre \citep{rodrigues2013learning}, Dog \citep{zhou2012learning}, and Rotten Tomatoes \citep{rodrigues2013learning}. Because now we have label matrix, we start from test dataset and item label matrix in Step 2 in Figure \ref{fig-exp}. Statistics of real-world datasets are shown in Table \ref{tab:dataset_real}. Due to extreme few labels given by workers, two settings are different from synthetic experiments. First, only workers providing more than $40\%$ labels are modeled and depth of decision trees is set to be $100$ in order to ensure quality of surrogate labels. Second, worker sampling is conducted over existing labels given by worker, and the sampling rate goes from $0.1$ to $0.5$ with an interval of $0.1$.

\begin{table}[!htbp]
\centering
	\caption{Performances of compared algorithms on real world datasets. Based on Wald's test, statistically better results are shown in \textbf{bold} fonts.}\label{tab:real}
	\begin{tabular}{cccccc}
	\noalign{\smallskip} \hline \noalign{\smallskip}
	\textbf{Dataset} & $\pi$ & \multicolumn{4}{c}{\textbf{Algorithms and Performances}}\\
	\noalign{\smallskip} \hline \noalign{\smallskip}
\multirow{8}{*}{Music} &  & \multicolumn{2}{c}{\textbf{DM}} & \multicolumn{2}{c}{\textbf{CI}}         \\ 
& 0.0                  & \multicolumn{2}{c}{$0.2286\pm0.0134$}   & \multicolumn{2}{c}{$\mathbf{0.2734}\pm0.0124$}            \\ 
& & \textbf{MV}       & \textbf{DRC-MV}      & \textbf{DS} & \textbf{DRC-DS} \\ 
&0.1                  & $0.2557\pm0.0087$         & $\mathbf{0.3449}\pm0.0097$            & $0.2166\pm0.0080$   & $\mathbf{0.2931}\pm0.0113$   \\ 
&0.2                  & $0.3610\pm0.0108$         & $\mathbf{0.4181}\pm0.0110$            & $0.3006\pm0.0102$   &$\mathbf{0.3300}\pm0.0075$        \\ 
&0.3                  & $0.4493\pm0.0105$         & $\mathbf{0.4824}\pm0.0088$            & $0.3559\pm0.0103$   &    $0.3716\pm0.0107$   \\ 
&0.4 & $0.5097\pm0.0091$                 & $0.5247\pm0.0073$            & $0.3926\pm0.0108$   & $0.3951\pm0.0104$       \\ 
&0.5                  & $0.5494\pm0.0123$         & $0.5601\pm0.0085$            & $0.4236\pm0.0127$   & $0.4159\pm0.0109$                           \\ \noalign{\smallskip} \hline \noalign{\smallskip}
\multirow{8}{*}{Dog} & & \multicolumn{2}{c}{\textbf{DM}} & \multicolumn{2}{c}{\textbf{CI}}         \\ 
&0.0                  & \multicolumn{2}{c}{$0.3637\pm0.0086$}   & \multicolumn{2}{c}{$\mathbf{0.4115}\pm0.0102$}            \\ 
& & \textbf{MV}       & \textbf{DRC-MV}      & \textbf{DS} & \textbf{DRC-DS} \\
&0.1                  & $0.5481\pm0.0092$         & $\mathbf{0.5860}\pm0.0068$            & $0.5703\pm0.0098$   & $\mathbf{0.5904}\pm0.0072$   \\ 
&0.2                  & $0.6644\pm0.0121$         & $\mathbf{0.6792}\pm0.0091$            & $0.6875\pm0.0080$   &$0.6919\pm0.0066$        \\ 
&0.3                  & $0.7238\pm0.0080$         & $0.7286\pm0.0066$            & $0.7450\pm0.0078$   &    $0.7431\pm0.0084$   \\
&0.4 & $0.7538\pm0.0073$                 & $0.7593\pm0.0074$            & $0.7786\pm0.0061$   & $0.7741\pm0.0077$       \\ 
&0.5                  & $0.7749\pm0.0064$         & $0.7746\pm0.0069$            & $0.7906\pm0.0069$   & $0.7872\pm0.0059$                           \\ \noalign{\smallskip} \hline \noalign{\smallskip}
\multirow{8}{*}{Tomato} & & \multicolumn{2}{c}{\textbf{DM}} & \multicolumn{2}{c}{\textbf{CI}}         \\ 
&0.0                  & \multicolumn{2}{c}{$0.5196\pm0.0041$}   & \multicolumn{2}{c}{$\mathbf{0.5373}\pm0.0036$}            \\ 
& & \textbf{MV}       & \textbf{DRC-MV}      & \textbf{DS} & \textbf{DRC-DS} \\ 
&0.1                  & $0.6263\pm0.0051$         & $\mathbf{0.6368}\pm0.0037$            & $0.6311\pm0.0033$   & $\mathbf{0.6425}\pm0.0037$   \\ 
&0.2                  & $0.7134\pm0.0034$         & $0.7169\pm0.0041$            & $0.7285\pm0.0035$   &$0.7329\pm0.0038$        \\
&0.3                  & $0.7637\pm0.0040$         & $0.7663\pm0.0039$            & $0.7925\pm0.0027$   &    $0.7924\pm0.0034$   \\
&0.4 & $0.8025\pm0.0030$                 & $0.8046\pm0.0031$            & $0.8374\pm0.0021$   & $0.8361\pm0.0026$       \\ 
&0.5                  & $0.8294\pm0.0036$         & $0.8295\pm0.0035$            & $0.8666\pm0.0030$   & $0.8629\pm0.0030$                           \\ \noalign{\smallskip} \hline \noalign{\smallskip}
\end{tabular}
\end{table}

For algorithm comparison, results are shown in Table \ref{tab:real}, which have four observations:
\begin{itemize}
    \item Given the same worker sampling rate, DRC-MV works better than MV and DRC-DS is better than DS, especially in low sampling rate cases. It shows our DRC approaches work in practice.
    \item As the sampling rate increasing, performances of DRC approaches converges to non-DRC approaches, which matches our understanding according to theorems.
    \item There are large improvements on Music and Dog datasets, while small improvements on Tomato datasets, potentially due to number of classes being too small. In other words, binary classification remains a easy task on Tomato dataset for workers. Bn contrast, it shows our DRC apporach is able to help in difficult multi-class classification problems.
    \item Compared between two zero cost methods, CI performs better than DM, which shows the effectiveness of supervised learners.
\end{itemize}

\subsubsection{Effectiveness of AIS and AWS}

\begin{figure}[!htbp]
	\centering
	\begin{minipage}{0.32\linewidth}\centering
		\includegraphics[width=\textwidth]{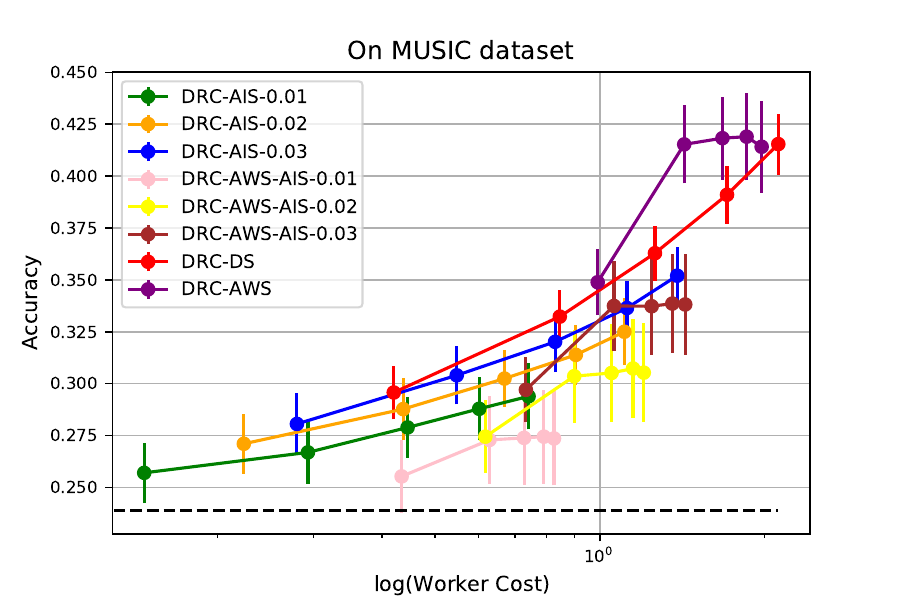}
	\end{minipage} 
	\begin{minipage}{0.32\linewidth}\centering
		\includegraphics[width=\textwidth]{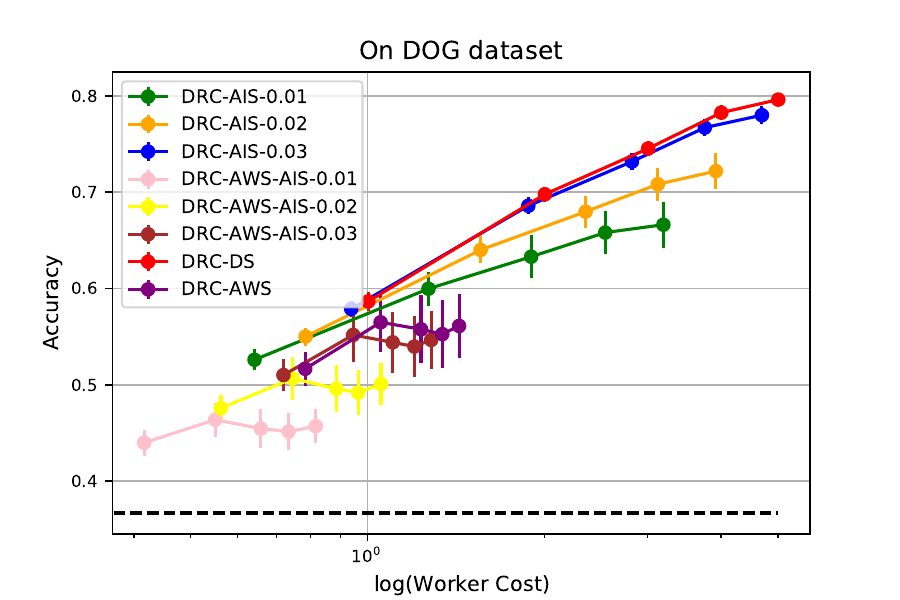}
	\end{minipage}
	\begin{minipage}{0.32\linewidth}\centering
		\includegraphics[width=\textwidth]{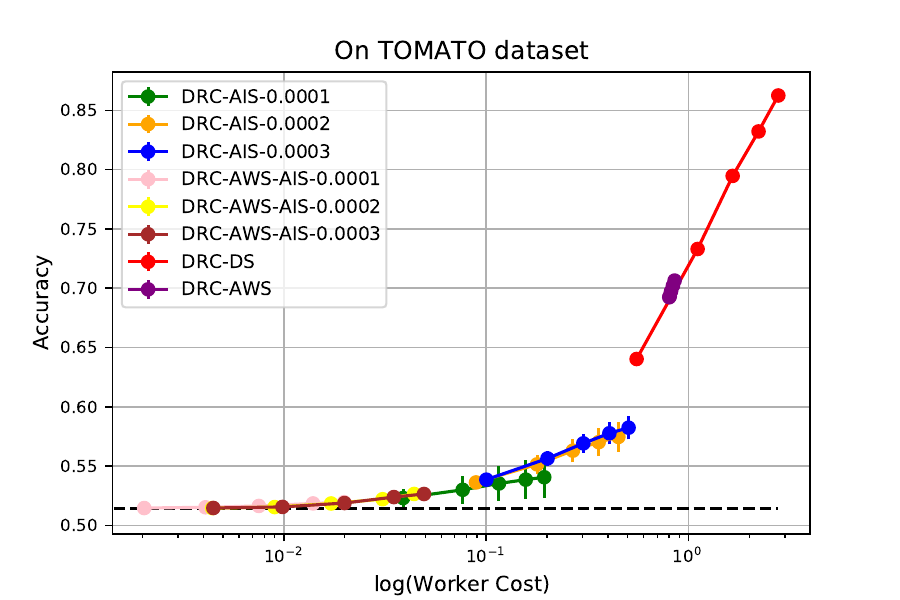}
	\end{minipage}
	\caption{Performances of DRC-AIS, DRC-AWS, DRC-AWS-AIS, and DRC-DS on real-world datasets.\label{fig-exp-real}}
\end{figure}

Similar to synthetic experiments, we do experiments of DRC-AWS-AIS, DRC-AIS, DRC-AWS, and DRC-DS on real-world datasets. $\lambda$ is set to be $1,2,4,7,10$ and $\rho$ is set in Figure \ref{fig-exp-real}. Other settings remain the same as above. 
There are four observations from results shown in Figure \ref{fig-exp-real}:
\begin{itemize}
    \item On Music dataset, DRC-AIS-0.01 enjoys the least worker cost, but with poor accuracy. DRC-AWS performs better than DRC-DS, both in cost and accuracy, which shows effectivenss of our weight-clipping technique.
    \item On Dog dataset, DRC-AWS-AIS enjoys the least worker cost, but performances increase quickly with DRC-AIS and DRC-DS.
    \item On Tomato dataset, it's hard to break the accuracy-cost tradeoff using AIS or ASW as the DRC-DS performance increases rapidly with more workers.
    \item There is an accuracy-cost tradeoff for all approaches on all datasets. For practical use, $\rho$ is suggested to set from $0.0001$ to $0.03$, small for easy tasks, for example, binary classification, while large for multi-class classification. $\lambda$ is suggested to set from $1$ to $c$, where $c$ is the average labels received per item. Small $\lambda$ leads to low cost and low accuracy while large $\lambda$ results in high cost and high accuracy.
\end{itemize}
    
\section{Conclusion}\label{sec:conc}
We formulate crowdsourcing as a statistical estimation problem and propose a new approach DRC to address it where worker imitation and doubly robust estimation are used. DRC can work with any base models such as Dawid-Skene model and majority voting and improve their performance. With adaptive item/worker selection, our proposed approaches are able to achieve nearly the same accuracy of using all workers but with less worker cost. In the future, there are many problems worth trying. Since item features are helpful for crowdsourcing problems, worker features can be taken into consideration as well. Also, if there are new workers joining the project, it needs special considerations.

\acks{The work is supported by an Adobe Data Science Award and a start-up grant made by the UCSB Department of Computer Science. The authors thank the anonymous reviewers and the associate editor for useful feedback.}

\vskip 0.2in
\bibliography{drc}
\bibliographystyle{plainnat}

\end{document}